\documentclass[copyright,creativecommons]{eptcs}
\providecommand{\event}{ICE 2010} % Name of the event you are submitting to
\usepackage{breakurl}             % Not needed if you use pdflatex only.

\usepackage{amsmath}
\usepackage{amssymb}
\usepackage{cases}
\usepackage{boxedminipage}

\title{A linear programming approach to general dataflow process network verification and dimensioning}
\author{Renaud Sirdey \qquad\qquad Pascal Aubry
\institute{CEA, LIST\\
Embedded Real-Time System Lab\\
91191 Gif-sur-Yvette Cedex, France}
\email{renaud.sirdey@cea.fr \quad\qquad p.aubry@cea.fr}
}
\def\titlerunning{A linear programming approach to general DPN verification and dimensioning}
\def\authorrunning{R. Sirdey \& P. Aubry}

\newcommand{\ZZ}{\mathbb{Z}}

\newcommand{\RR}{\mathbb{R}}

\newcommand{\qp}{\text{\textsf{qp}}}
\newcommand{\qc}{\text{\textsf{qc}}}

\newcommand{\zip}{z_{\text{IP}}}
\newcommand{\zlp}{z_{\text{LP}}}

\begin{document}
\maketitle

\begin{abstract}
In this paper, we present linear programming-based sufficient conditions, some of them polynomial-time, to establish the liveness and memory boundedness of general dataflow process networks. Furthermore, this approach can be used to obtain safe upper bounds on the size of the channel buffers of such a network.
\end{abstract}

\section{Introduction}

\par With the frequency version of Moore's law coming to an end, a new generation of massively multi-core microprocessors is emerging. This has triggered a regain of interest for the so-called dataflow programming models in which one expresses computation-intensive applications as networks of concurrent tasks interacting through (and only through) unidirectional FIFO channels. 

%These models are of the uttermost practical relevance because they relieve the programmer from one of the main pitfall of parallel programming|having to explicit complex synchronization schemes|as well as allow the natural expression of more than enough parallelism for chips in the few hundred cores range, at least in the (rather wide) domain of signal and image processing. 

%\par On top of mainstream compilation skills, compiling a dataflow program in order to achieve a high level of dependability and performance on such an architecture involves fairly advanced techniques mostly in terms of scarce and interrelated resources allocation (which requires solving a whole bunch of discrete optimization problems) as well as formal analysis of computer programs (so as, in particular, to guarantee certain liveness properties). This paper deals with the latter of those two aspects.

\par Our main result is a linear programming model (Sect. \ref{sec:model}) of the states of a \emph{general} Dataflow Process Network (DPN), in the sense of Lee \& Parks \cite{LEE-95}, which allows to obtain a polynomial-time sufficient condition for both liveness (in a sense which is defined in Sect. \ref{sec:prop}) and memory boundedness (Sect. \ref{sec:dim}). Furthermore, this approach can be turned into a safe buffer dimensioning method.

%\par It should straightaway be emphasized that the DPN formalism must not be confused with the much more (although better known) restrictive Synchronous Data Flow (SDF) model of Lee \& Messerschmitt \cite{LEE-87} or the Cyclo-Static Data Flow (CSDF) model of Bilsen et al. \cite{BIL-96}. In essence, the DPN formalism has the same power as the Kahn Process Network formalism and, as such, certain properties such as memory boundedness are not decidable whereas the aforementioned two models (as well as most of their variants) are benign with that respect. 

%\par The present paper is organized as follows. Sect. \ref{sec:model} presents our integer linear model of the states of a DPN. Sect. \ref{sec:prop} defines the liveness property in which we are interested and shows how to verify it by proving that certain integer linear systems are inconsistent. In Sect. \ref{sec:dim}, we introduce an integer linear program which solution gives a sufficient condition for establishing both liveness and memory boundedness as well as demonstrate that an equivalent (though polynomial-time) condition can be obtained by considering its continuous relaxation. Lastly, Sect. \ref{sec:algo} deals with algorithmic considerations.

\section{Modelling system states}
\label{sec:model}

\subsection{Notations and general assumptions}

\par Let $T$ and $F$ respectively denote the set of tasks and channels.

\par To each task $t\in T$, we associate a state-transition graph $G_t=(\{v_t^{(0)}\}\cup V_t,\{\tau_t^{(0)}\}\cup A_t)$ (parallel arcs and loops are allowed), where $v_t^{(0)}$ denotes the initial state of task $t$ and where $\tau_t^{(0)}$ denotes the initial transition of that task, this transition being unique and unconditional. Also, given $t\in T$, $P_t\subseteq F$ (respectively $C_t\subseteq F$) denotes the set of channels in which $t$ produces (respectively consumes) data. Note that we have $P_t\cap C_t=\emptyset$. Also note that for each $t,t'\in T^2$, $t\neq t'$, we have $P_t\cap P_{t'}=\emptyset$ as well as $C_t\cap C_{t'}=\emptyset$.

\par Let $t\in T$, $\tau\in A_t$ and $f\in P_t$ (respectively $f\in C_t$), $\qp_{\tau f}$ (respectively $\qc_{\tau f}$) denotes the amount of data produced (respectively consumed) in channel $f$ by task $t$ when transition $\tau$ is executed. An additional constraint is that, $\forall t\in T$, $\forall \tau\in A_t$,
\begin{equation}
\label{eq:effect}
\sum_{f\in P_t}\qp_{\tau f}+\sum_{f\in C_t}\qp_{\tau f}>0,
\end{equation}
thereby excluding the existence of transitions having no effect on any of the channels.

\par Given $f\in F$, $p_f$ and $c_f$ respectively denote the tasks which produce and consume data in channel $f$. Also, $d_f$ denote the capacity of the buffer associated to channel $f$ (depending on the problem at hand $d_f$ can be either given or unknown, as we shall later see).

\par In the sequel, it is further assumed without loss of generality that the network graph, the directed graph having the tasks as vertices and the channels as arcs, is (simply) connected.

\subsection{Variables and linear constraints}

\par For all $\tau\in\bigcup_{t\in T}\{\tau_t^{(0)}\}\cup A_t$, we introduce a variable denoted by $n_\tau\in\ZZ^+$ which indicates the number of times transition $\tau$ has been executed. In order for the $n_\tau$ to represent admissible system states, a number of (linear) constraints must be satisfied.

\par \emph{Initialization constraints}. Let $t\in T$, initial transition $\tau_t^{(0)}$ must have been executed once and only once, thus, $n_{\tau_t^{(0)}}=1$.

\par\emph{Conservation constraints}. Let $t\in T$ and $v\in V_t$. We then have the following constraint: %\footnote{As usual, $\omega^-_{G_t}(v)$ (respectively $\omega^+_{G_t}(v)$) denotes the set of arcs in $\{\tau_t^{(0)}\}\cup A_t$ which head (respectively tail) is $v$. We will however use the lighter notations $\omega^-(v)$ and $\omega^+(v)$ when there is no ambiguity on the graph which is referred to.}:
$$
\sum_{\tau\in\omega^-(v)}n_\tau-1\leq\sum_{\tau\in\omega^+(v)}n_\tau\leq\sum_{\tau\in\omega^-(v)}n_\tau.
$$
Such a constraint simply reflects the fact that for the $n_\tau$'s to represent an admissible system state, vertex $v$ must have been left as many times it has been entered or as many times minus one (in which case it defines the current state of $t$). If $\sum_{\tau\in\omega^+(v)}n_\tau=\sum_{\tau\in\omega^-(v)}n_\tau-1$, then the system state described by the $n_\tau$'s is such that task $t$ is in state $v$.

\par For convenience, let $\gamma_v=\sum_{\tau\in\omega^-(v)}n_\tau-\sum_{\tau\in\omega^+(v)}n_\tau$ meaning that $\gamma_v=0$ if task $t$ is not in state $v$ and $1$ otherwise. Remark that state $v_t^{(0)}$ is duly excluded from the previous sums as this state is by definition left once and never entered.

\par\emph{Unicity constraints}. Furthermore, for the system state described by the $n_\tau$'s to be admissible, each task must be in one and only one state. That is, for each $t\in T$, if $\sum_{v\in V_t}\gamma_v=1$. Again, note that $v_t^{(0)}$ is duly excluded from the previous sum.

\par\emph{Consistency constraints}. Let $f\in F$, let $\qp_f=\sum_{\tau\in A_{p_f}}n_\tau\qp_{\tau f}$ and $\qc_f=\sum_{\tau\in A_{c_f}}n_\tau\qc_{\tau f}$ respectively denote the amount of data so far produced and consumed on channel $f$ (for convenience). For the $n_\tau$'s to describe a valid system state, we must have $\qp_f\geq\qc_f$.

\par\emph{Capacity constraints}. Also, for each $f\in F$, we must have
\begin{equation}
\label{eq:cap}
\qp_f-\qc_f\leq d_f.
\end{equation}

% \par Needless to emphasize that although the solutions of that system define what we refer to as admissible system states, some of these states may not be reachable.

\section{Modelling undesirable system properties}
\label{sec:prop}

\subsection{Strong and weak blockedness}

\par In a given system state, a task $t\in T$ is \emph{strongly blocked} when it is in a state $v$ in which no outgoing transition can be executed. Consider the following sets of constaints, for each $\tau=(v,v')\in A_t$,
\begin{numcases}{}
\label{eq:prereq:1} \gamma_v\leq0, \\
\label{eq:prereq:2} \qp_f-\qc_f\leq\qc_{\tau f}-1, & for each $f\in C_t$,\\
\label{eq:prereq:3} \qc_f-\qp_f\leq\qp_{\tau f}-d_f-1, & for each $f\in P_t$.
\end{numcases}
Then, strong blockedness means that for each $\tau=(v,v')\in A_t$ either constraint \eqref{eq:prereq:1} or at least one the constraints of type \eqref{eq:prereq:2} or at least one of the constraints of type \eqref{eq:prereq:3} is satistifed for task $t$.

\par In a given system state, a task $t\in T$ is \emph{weakly blocked} when it is in a state $v$ in which not all outgoing transitions can be executed. Consider the following sets of constraints, for each $t\in T$ and for each $v\in V_t$,
\begin{numcases}{}
\label{eq:dang:1} \gamma_v\leq0, \\
\label{eq:dang:2} \qp_f-\qc_f\leq\qc_{\tau f}-1, & for each $\tau=(v,v')\in A_t$ and for each $f\in C_t$,\\
\label{eq:dang:3} \qc_f-\qp_f\leq\qp_{\tau f}-d_f-1, & for each $\tau=(v,v')\in A_t$ and for each $f\in P_t$.
\end{numcases}
Then, weak blockedness signifies that for each $t\in T$ and for each $v\in V_t$ either constraint \eqref{eq:dang:1} or at least one the constraints of type \eqref{eq:dang:2} or at least one of the constraints of type \eqref{eq:dang:3} is satistifed for task $t$.

\par The above strong blockedness property is suitable to model non deterministic tasks whereas the weak blockedness property allows modelling deterministic ones (enforcing the fact that, when the task is in a given state, all outgoing transitions are feasible so as to guarantee that the next transition the task \emph{has} to do is possible).

\subsection{Sufficient liveness conditions}

\par Although the definitions of the strong and weak blockedness properties involves disjunctive constraints, these can be linearized using standard linear programming modelling techniques (see e.g., \cite{NEM-99}).

\par Thus, given a DPN and a dimensioning $d\in\ZZ^{|F|}$ we have shown how to formulate an integer linear system of inequalities,
\begin{equation}
\label{eq:int:lin:sys}
\{x\in\ZZ^n:Ax\leq b(d)\}
\end{equation}
which \emph{inconsistency}, i.e. $\{x:Ax\leq b(d),x\in\ZZ^n\}=\emptyset$, is \emph{sufficient} to establish the liveness of the network. A fortiori, the inconsistency of the continuous relaxation of that system, $\{x:Ax\leq b(d),x\in\RR^n\}=\emptyset$, provides a polynomial-time (weaker) sufficient condition to establish that property. 

\section{Memory boundedness of a DPN}
\label{sec:dim}

\subsection{Monotony with respect to dimensioning}
\label{sec:monotony}

\par As shown by Lee \& Parks \cite{LEE-95} the DPN formalism is equivalent to the well-known Kahn Process Networks (KPN) formalism. Thus, DPN also exhibit the determinism property exhibited by KPN whereby, for a given input, the data circulating on the channels does not depend on the execution trace. This very convenient property allows, still for a given input, to derive general network properties from properties exhibited by particular traces of execution.

\par The KPN formalism assumes blocking reads and non blocking writes, which may induce an infinite memory requirement on some channels. However, a KPN $K$ subject to capacity constraints on its channels can straightforwardly be turned into another KPN $K(d)$ ($d\in\ZZ^{+n}$): all is needed is to emulate a blocking write with a blocking read e.g., to mirror each channel $f$ by an opposite channel $f'$ initially provided with $d_f$ data and to require that each write, respectively read, operation on $f$ be mirrored by a equivalent read, respectively write, operation on $f'$. Of course, the properties of KPN $K$ with respect to the data circulating on the channels are not preserved by this transformation and, in particular, $K(d)$ may not be deadlock-free despite of the fact that $K$ is.

\par In essence, the liveness property defined in the previous section ensures that, \emph{for all possible inputs}, an infinite amount of data circulates on \emph{at least one} channel. Assume that a KPN $K(d)$ is live, then a straightforward consequence of the determinism property of KPN is that any KPN $K(d')$ such that $d'\geq d$ (i.e., $\forall f\in F$, $d'_f\geq d_f$) is also live. Indeed, the liveness of $K(d)$ implies that, for any input $\alpha$, any given trace of execution $\omega(\alpha)$ of $K(d)$ is such that an infinite amount of data circulates on at least one channel and since $\omega(\alpha)$ is also a valid trace of execution of $K(d')$, from the determinism property, all traces of execution of $K(d')$ on $\alpha$ are also such that an infinite amount of data circulates on at least one channel and, thus, $K(d')$ is live. It follows that, given a DPN, if we can find $d\in\ZZ^{|F|}$ such that either $\{x:Ax\leq b(d),x\in\ZZ^n\}=\emptyset$ or $\{x:Ax\leq b(d),x\in\RR^n\}=\emptyset$, then for any $d'\in\ZZ^{|F|}$ such that $d'\geq d$, the DPN is live.

\subsection{Verifying memory boundedness}

\par Recall integer linear system \eqref{eq:int:lin:sys}, assume that $d_f=z$ ($\forall f\in F$) and consider the following Integer Linear Program
\begin{numcases}{}
\nonumber \zip = \text{Maximize}~z\\
\nonumber A'y\leq b', \\
\label{eq:int:lin:prog} y\in\ZZ^{n+1},
\end{numcases}
where vector $y$ is the concatenation of vector $x$ and scalar $z$. This program is straightforwardly derived from system \eqref{eq:int:lin:sys} by replacing $d_f$ by $z$ in constraints \eqref{eq:cap} as well as constraints \eqref{eq:prereq:3} or \eqref{eq:dang:3} (depending on which of the two applies) and by moving $z$ to the LHS.

\par In essence, from the monotony property derived in Sect. \ref{sec:monotony}, any solution to the above program provides \emph{the largest value of $z$ such that the network is not live} and, thus, for any dimensionning $d\in\ZZ^{|F|}$ such that $\forall f\in F$, $d_f\geq \zip$ the network is live. Three cases can then occurs. Case 1: the ILP has no solution, a degenerate case which can occur only for networks with no channels (since, due to Eq. \eqref{eq:effect}, all transitions are effective). This latter case in hereafter ignored. Case 2: $\zip<\infty$, which is sufficient to establish that the network is live and memory bounded. Case 3: $\zip=\infty$ in which case we cannot conclude with respect to both liveness and memory boundedness. Furthermore, when $\zip<\infty$, letting $d_f=z_{\text{IP}}+1$ ($\forall f\in F$) gives a (presumably small) channel buffer dimensioning which guarantees liveness.

\par Again, it is possible to consider the continuous relaxation $\zlp$ of program \eqref{eq:int:lin:prog}. In particular, when one wishes only to determine wether or not $\zip<\infty$ then it is necessary and sufficient to determine whether or not $\zlp<\infty$ as $\zip<\infty\Leftrightarrow\zlp<\infty$. Indeed, a well known fact in the theory of linear and integer programming \cite{NEM-99} states that if the integer hull $P_I$ of a rational polyhedron $P$ (i.e., the convex hull of the integral vectors in $P$) is nonempty then $\max\{cx:x\in P\}$ is bounded \emph{if and only if} $\max\{cx:x\in P_I\}$ is bounded and, provided that $0\in\{y:A'y\leq b',y\in\ZZ^{n+1}\}$, this result applies to program \eqref{eq:int:lin:prog} and its relaxation. It follows that $\zlp<\infty$ provides a polynomial-time sufficient condition to establish both liveness and boundedness of a DPN which is equivalent to $\zip<\infty$.

\section{Remarks on algorithmic aspects}
\label{sec:algo}

\par Although $\zlp$ can be computed in polynomial-time, it can be expected, when $\zlp<\infty$, that the integrality gap, $\zlp-\zip$, is often quite large. Thus, should one wishes to obtain a tight (if not the tightest) upper bound on the channel buffer dimensioning, there is a practically relevant need to either solve program \eqref{eq:int:lin:prog} or at least to decrease the aforementioned integrality gap.

\par Regarding the resolution of program \eqref{eq:int:lin:prog}, it should be emphasized that the polyhedron $P_I=\{y:A'y\leq b',y\in\ZZ^{n+1}\}$ (recall that the integer hull of a polyhedron is also a polyhedron) is generally not a polytope (since the $n_\tau$ are not necessarily bounded). Therefore, in the general case, procedures which enumerate integer points inside the polyhedron, as those used in most off-the-shelf integer linear programming solvers, are doomed not to terminate.

\par Thus, in order to guarantee termination in the present context, an (exterior) cutting plane approach must be used. As an example, Gomory's cutting plane algorithm is guaranteed to terminate (though generally after a prohibitively long time). A more practically promising approach, consists in using specially tailored classes of inequalities derived from a polyhedral study of the geometric structure of $P_I$ so as to derive a custom cutting plane algorithm. %Although such an algorithm should not be expected to always provide an optimal integral solution, it can be expected to allow to significantly decrease the integrality gap much faster than an general purpose cutting plane algorithm (see e.g., \cite{NEM-99}).

%\section{Conclusion}

%\par We have presented a linear programming model for the states of general dataflow process networks and shown how it could be used to derive a polynomial-time sufficient condition to establish the liveness and memory boundedness of such networks. Although this approach can also be used for obtaining a safe upper bound on the channel buffer sizes by means of (continuous) linear programming, finding a tightest upper bound requires solving an integer linear program with the particularity that the associated polyhedron is not a polytope.

%\par Further work needs to investigate the geometric structure of this polyhedron so as to define classes of facet-defining inequalities in order to design an approximate resolution cutting plane algorithm. Still, it is expected that the present paper contributes in paving the way towards the application of the powerful algorithmic techniques which have been developed in the field of Combinatorial Optimization to the verification of certain classes of process networks of practical relevance for the presently emerging new generation of massively multicore microprocessors.

\bibliographystyle{eptcs} % or whatever you prefer

\end{document}